\\
\documentstyle[12pt,aasms]{article}
\received{\hrule}
\accepted{\hrule}
\journalid{337}{15 January 1989}
\articleid{11}{14}
\def\l{$\lambda$}

\def\msol{$M_\odot$}

\def\ltsima{$\; \buildrel < \over \sim \;$}
\def\simlt{\lower.5ex\hbox{\ltsima}}            
\def\gtsima{$\; \buildrel > \over \sim \;$}
\def\simgt{\lower.5ex\hbox{\gtsima}}            

\def\h{$h^{-1}$}
\def\a{$\alpha$}
\def\b{$\beta$}
\def\kms{km~s$^{-1}$}
\begin{document}

\title{\sc Multiple High-Velocity Emission-Line Systems in the E+S Pair
CPG 29\altaffilmark{1}}

\author{P. Marziani and W. C. Keel}
\affil{Department of Physics \&\ Astronomy, University of Alabama,
Tuscaloosa, AL 35487--0324}

\author{D. Dultzin--Hacyan}
\affil{Instituto de Astronomia, UNAM, Apartado Postal 70--264,
M{\'e}xico D. F., M{\'e}xico}

\author{J. W. Sulentic}
\affil{Department of Physics \&\ Astronomy, University of Alabama,
Tuscaloosa, AL 35487--0324}

\altaffiltext{1}{Based in part on data collected at the European Southern
Observatory, La Silla, Chile;  Kitt Peak National Observatory,
National Optical Astronomy
Observatories, operated by AURA under cooperative agreement with the NSF; at
Lowell Observatory, Flagstaff, Arizona; Observatorio Astronomico Nacional,
San Pedro Martir, Baja California, Mexico; and
Special Astrophysical Observatory, formerly USSR Academy of Sciences.}

\begin{flushleft}
\begin{abstract}

A detailed study of the mixed-morphology galaxy pair CPG 29 (Arp 119,
VV 347) shows spectacular spectroscopic peculiarities in the southern
(spiral) component (Mkn 984) including a spatially resolved region,
roughly aligned along the minor axis of the galaxy, with
multiple emission-line components redshifted by as much as 1300 km
s$^{-1}$ with respect to the LINER nucleus.  Strong [OI]\l6300 and
[SII]\l\l6716,6731 emission suggest shock ionization. The rest of the galaxy's
disk is spectroscopically undistinguished, with emission lines characteristic
of disk H II regions and, in large part, ordered rotation.

We discuss the following four possible explanations for the morphological and
spectroscopic
peculiarities in the spiral: (1)  a super--wind scenario,
supported by the high FIR luminosity of the spiral, and by
 emission line activity  similar to that observed in
FIR strong galaxies; (2) motion of the spiral through a hot, relatively
dense intergalactic medium, possibly associated with the elliptical,
that could affect the star formation rate and create a brightened disk rim;
(3) a faint companion in direct collision with the disk, at
$\Delta v_r \sim 1000$ \kms, that would straightforwardly explain the
existence of the different redshift systems; (4) a pole on collision by the
elliptical, that could produce the disturbed morphology and other properties of
the spiral.

The elliptical/spiral pole--on collision explains the high velocity line
emitting regions as tidally stripped gas, and accounts for the presence of gas
ionized by moterate velocity shocks. The morphology of Mrk 984 is consistent
with a ring galaxy and numerical simulation suggest that the elliptical can
emerge unscathed from a relatively high velocity crossing.

\end{abstract}

\keywords{Galaxies: Interactions -- Galaxies: Nuclei --
Galaxies: Active -- Galaxies: Evolution -- Galaxies: Individual (Mrk 984,
CPG 29)}

\newpage

\section{Introduction}
Gravitational interaction between galaxies is responsible for a wide
variety of phenomena. Many morphological peculiarities in spiral
galaxies, such as bars, tails, and external rings, have been related to
tidal effects (Toomre \&\ Toomre 1972; Lynds \&\ Toomre 1976;
Elmegreen, Elmegreen, \&\ Bellin 1991).  Comparisons of interacting and
isolated galaxy samples suggest that spectroscopic and photometric
properties are also modified in interacting systems.  Galaxies in the
Arp (1966) atlas have excess radio continuum emission (Sulentic 1976)
and a wider dispersion in UBV colors than isolated galaxies (Larson
\&\ Tinsley 1978). H\a\ luminosities in disks and nuclei of interacting
systems are on average much higher ($\times$10) than in isolated
galaxies (Kennicutt et al. 1987; Bushouse 1987).
Far-IR emission is  also strongly enhanced in interacting galaxies
(Bushouse 1987; Xu \&\ Sulentic 1991).  All these results can be linked
to an enhanced star formation rate (SFR) in interacting galaxies.

Gravitational interaction may also be the ultimate triggering mechanism
of non-thermal nuclear activity. Dahari (1985), and Rafanelli
\&\ Marziani (1992a) found a significant excess of interacting systems
among Seyfert galaxies. It is still unclear how interaction may
ultimately lead to the production of an active nucleus. An evolutionary
link between a starburst and the origin of an  active nucleus  is
fashionable (either remnants of evolved massive  stars may coalesce to
form a massive black hole (e. g.  Weedman 1983), or  mass  loss may
provide accreting  matter,  if the black hole was formed earlier,
Norman \&\ Scoville 1988; Heckman, Armus \&\ Miley 1990).

Object number 29 in the Catalog of Isolated Pairs of Galaxies (CPG 29;
Karachentsev 1972), also known as VV 347 and Arp 119, is a
Spiral+Elliptical pair (components hereafter referred to as CPG 29S and
CPG 29N or {\bf S} and {\bf E} for brevity) of galaxies, with projected
component separation of
approximately 1 arcmin.  The southern component (0116+121$\equiv$ CPG
29S$\equiv$ Mrk
984 $\equiv$ UGC 849) has been classified as type Sdm pec. in the RC3
(de Vaucouleurs et al. 1991).  The peculiarities in {\bf S} are not
confined to morphology, but include an unusual emission-line spectrum.
The broad and complex line profiles of this object were first noted
by Osterbrock \&\ Dahari (1983).
{\bf S} has been catalogued as a Seyfert 2 galaxy (Artyuk, et al., 1982),
but the nuclear ionization level is too low for it to be considered as
a {\em classical} Seyfert 2 (Osterbrock \&\ Dahari 1983). The line ratios are,
at first
glance, more typical of a LINER (Dahari \&\ De Robertis 1988;
Mazzarella \&\ Boroson 1993)

The most
puzzling aspect presented in this paper results from long slit spectroscopy
where we find (at
least) four redshift systems covering a range $\Delta v_r \approx$1300
\kms\ in excess of the radial velocity of the underlying galaxy.
Emission  in the high velcoity systems occurs in a region roughly aligned close
to the minor axis of the galaxy, associated with the nucleus of Mrk 984, and a
northern blob (see Fig. 1).

 In
\S \ref{sec:obs} we describe new imaging and spectroscopic
observations, reduction, and data analysis. Morphology, velocity curves and
properties of the line emitting regions are presented in \S\ 3. In \S\ 4 we
analyze the ionization mechanism, estimate the galactic masses, and suggest a
hint to  the star
formation history for {\bf S}. In \S\ 5 we
discuss four possible explanations for this remarkable system.

\section{Observations and Data Reduction}
\label{sec:obs}
\subsection{Imaging}
The first part of Table 1 summarizes the imaging observations.
The KPNO B and V images
were obtained as part of a survey of mixed pairs by N. Sharp and
J. W. Sulentic. Images in the red were obtained
at Lowell Observatory. A filter
of FWHM 75 \AA\ centered near 6900 \AA\ traced  H$\alpha$
and [N II] line emission. An emission--line image was produced by subtraction
of a scaled R image, instead of pure
continuum, giving the right emission-line morphology but underestimating
the flux by a known factor, given the ratio of the two bandwidths.

\subsection{Long-slit Spectroscopy}
A journal of spectroscopic observations is given in the second part of Table 1.
The 2.1 m (F/7.5) telescope of the Mexican Observatorio Astronomico Nacional
located on the San Pedro Martir Sierra (Baja California, Mexico) was equipped
with a Boller \&\ Chivens spectrograph and a 600 l/mm grating (2.5 arcsec
slit). The KPNO 2.1 m telescope was used with the Gold Camera spectrograph (2.5
arsec slit).  The ESO 1.5m telescope was equipped with a Boller \&\ Chivens
spectrograph and 400 l/mm grating.

All slit spectra were reduced in essentially the same way. Frames were bias
subtracted and flat field corrected. Wavelength calibration, from Helium-Argon
spectra obtained after each
exposure, gave an rms precision $\le$ 0.05 \AA. Spectral resolution (FWHM) was
3.8
\AA\ (KPNO), $\approx 5$ \AA\ (SPM), and $\approx$4 \AA\ (ESO).
The spectra were rebinned to linear wavelength scales and corrected for
cross-dispersion distortion, and finally extinction corrected and flux
calibrated. Standard stars were not observed on the same nights as
the CPG 29 data at ESO and KPNO, so the response curve was
computed from stars observed on different nights of the same
runs. We can be confident that large errors in the spectrophotometry are
absent, since the continuum
fluxes of the 1990 and 1992 KPNO spectra, and of the ESO spectra for {\bf A}
and
{\bf B}, agree well. The accuracy in the absolute continuum fluxes
at 6500 \AA\ is estimated to be $\sim 20$ \%. Continuum fluxes for A (P.A.
=7$^\circ$) obtained at SPM are about 25 \%\ lower than those obtained from the
KPNO and ESO spectra.
We did not scale the SPM fluxes (reported in Table 2 \&\ 3) to the average
continuum
value. The reader must be aware that an underestimate of the flux as large as
25 \%\ is possible for the emitting regions on the SPM spectra reported in
Tables 2 \&\ 3 (P. A. = 116$^\circ$).

Some emission line fluxes disagree markedly (see Table 2), most probably due to
the peculiar extended emission in this object.  The flux of the H\a\
+[NII]\l\l6548,6583 blend in the circumnuclear region is comparable to that of
the nucleus. A slight misplacement or a change of orientation or atmospheric
seeing might change the emission-line fluxes appreciably. This is especially
true for fluxes of the extranuclear regions closest to {\bf A}. The uncertainty
of the
line fluxes measured in these region is probably larger than that
associated with the other nuclear emitting regions, and is estimated to
be around 40 \%.

\subsection {Area Spectroscopy}

Part of {\bf S} was observed in the region 6200-7000 \AA\ using
the multipupil fiber spectrograph
(MPFS) at the 6-meter telescope (BTA) of the Special Astrophysical
Observatory (SAO) of what was then the USSR
Academy of Sciences. This system, described by Afanasiev, Vlasiuk, Dodonov \&\
Sil'chenko (1990), uses a lens-coupled fiber array to simultaneously measure
the spectra of an $8 \times 12$ array of 1.25-arcsecond square
apertures. The detector was a $512 \times 512$ TV-based photon-
counting system. Monochromatic maps of the continuum and line features were
produced.

\section{Results}

\subsection{Morphology}
\label{sec:morph}

The nuclei of {\bf S} and {\bf E} are separated by 54"$\pm$ 0".2, or a
projected linear separation of 38 h$^{-1}$ kpc (for $H_0 = 100\times h$
km s$^{-1}$ Mpc$^{-1}$ so that 1" $\approx$ 688\h\  pc; see Fig.
\ref{fig:vimage}. Markarian 983  is located approximately 2'.5 south of {\bf
S}.  It shows a heliocentric $v_r$\ $\approx$ 14,400 km
s$^{-1}$; Palumbo, Tanzella-Nitti \&\ Vettolani 1983), consistent with a common
distance for
this galaxy and CPG 29. Mkn 983 has been described as having multiple
nuclei or hot spots (Mazzarella, Bothun, \&\ Boroson  1991; Mazzarella \&\
Boroson
1993).  It is within the field of our Lowell images, and in $R$ shows
what may be face-on spiral structure with a pronounced bright knot
along an armlike feature west of the nucleus. This knot is by far the
dominant source of H$\alpha$ emission in Mkn 983, with weaker emission
at the nucleus and along the apparent western arm.

The spiral component subtends $\approx$ 72" $\times$32".  ($\approx$ 49
$\times$ 22 \h\ kpc).The southeastern part of {\bf S} shows fairly
regular spiral structure, typical for Sc galaxies. Our new images
suggests an Sc peculiar type for {\bf S} The axial ratio is consistent
with an inclination (i$\sim 63^\circ$ (the UGC values imply
i=66$^\circ$). Direct estimation of the inclination from the axial
ratio may be misleading because of either warping of the disk or a
tilted outer ring (see \S \ref{sec:ring}).  A most relevant {\it morphological}
peculiarity in {\bf S} is the presence of two knots to
the north of {\bf A} and towards the nucleus of {\bf E} -- knot {\bf B}
($m_{B} \approx 20.0$, d$_{BA} \approx 6"$) and knot {\bf C} ($m_{B}
\approx$ 21.0, Mazzarella \&\ Balzano [1993], $d_{AC} \approx 13".6$; Fig.
\ref{fig:vimage}).
{\bf B} and {\bf C} are located approximately along the
separation vector of the pair. {\bf B}
appears starlike on the blue
frame (see Fig. \ref{fig:contour})  and becomes distinctly elongated on the V
image (Fig. \ref{fig:vimage}).
At least two, and more likely three, ring--like structures might be present in
{\bf S}. An inner ring surrounds the nucleus {\bf A} at r= 6-7". Knot {\bf B}
is
superimposed on this feature but the ring itself is not a site of
strong H\a\ emission (see Figure \ref{fig:contour}). The bright arc {\bf R} is
a part of a second ring at d$\approx$ 14" from the nucleus. A line of faint
knots,
probably HII regions, that skirts the southern edge of the disk,  and the arc
on the north-east visible on Fig. \ref{fig:vimage}, could be a third ring--like
feature (\S\ \ref{sec:ring}).

Figure \ref{fig:contour} compares the inner regions of {\bf S} on the H\a\ and
B
images.  Two intriguing results emerge: (1) the nucleus ({\bf A}) on
the blue image lies between two knots of H\a\ emission, but close to the
(fainter) southern one.  A further look at the blue image shows that {\bf A}
is partially resolved and slightly fainter on the north east side. (2)
The blue image shows a knot towards the east of the nucleus that has no
obvious counterpart on the H\a\ image. Conversely, the H\a\ images
shows emission to the south without obvious counterpart in the blue.

Geometric and luminosity profiles for {\bf E} are shown in Figure
\ref{fig:ell}\ for
both the B and V images. There is no evidence for variation in
the isophotal ellipticity (middle panel  of Fig. \ref{fig:ell}) outside of the
2"
seeing disk. The major-axis orientation (upper panel of \ref{fig:ell}) shows a
small rotation from $\approx 14^\circ$ down to $\approx 8^\circ$
at $d \approx 15$",  with a global change of $\mid\Delta $ P.A.
$\mid\simlt 10^\circ$. The surface-brightness profile (given in arbitrary units
in the bottom panel of Fig. \ref{fig:ell}) shows minor deviations from an
$r^{-1/4}$ profile including a small ripple at
d$\approx$ 16", followed by a steepening outwards. The isophotal
centroid is constant with intensity to within $\approx 1".5$.

\subsection{Definition of the Line Emitting Regions}

Line emitting regions were isolated to match the most prominent
morphological features observed in the B and V images. The definitions
are based  on a careful analysis of the cross-dispersion profile of the
H\a\ + [NII]\l\l 6548,6583 blend, and the continuum at $\approx$ 6500
\AA.  The definition of each emitting region can be found on Fig.
\ref{fig:pa8},
 \ref{fig:pa116}, \ref{fig:pa81}.
The heliocentric $v_r$
for each region was estimated from several lines among [NII]\l6583,
H\a, H\b, [OIII]\l5007, [OI]\l6300.  Blending prevented us from
measuring some of the lines in particular regions.  Heliocentric radial
velocities estimated from different lines are the same within the
observational uncertainties for all components over all the regions we
identified. In particular, the $v_r$\ estimated from [OIII]\l\l
4959,5007 is in agreement with $v_r$\ estimated from the Balmer lines.
Table 3 reports fluxes \&\ widths
of regions having single peaked emission line profiles.

\subsection{Extended Emission}
\label{sec:spec}

The most striking feature in the spectra taken at P.A. =8$^\circ$ are
the apparently broad (FWHM  $\sim$ 1200 \kms), multi--peaked emission
line profiles. Broad profiles are observed for [OI]$\lambda\lambda$6300, 6363;
[OIII]$\lambda$5007; H$\beta$, H$\alpha$, [NII]$\lambda\lambda$6548, 6583 and
[SII]$\lambda\lambda$6716, 6731. They are seen in {\bf A} and extend
north-south for approximately 10 arcsec (regions {\bf A}, {\bf B}, and
{\bf D}). This extended region appears in the spectra at P.A. = 3$^\circ$ and
8$^\circ$; see Fig. \ref{fig:pa8}). At P.A.=116$^\circ$, broad line profiles
are observed only from {\bf A} and in
the region immediately east of the nucleus (region {\bf F}; Fig.
\ref{fig:pa116}).  The spectra
taken at P.A. = 81$^\circ$\ (Fig. \ref{fig:pa81}) show multi--peaked profiles
only at knot
{\bf B}.

The profile variation of the blend H\a+[NII]\l\l6548,6583 along the slit at PA=
8$^{\circ}$ is shown in
Fig. \ref{fig:blend}. Analysis of the H\a\ + NII]\l\l6548,6583 profile in {\bf
A} and in the
surrounding emission region shows that each line of the blend is made
up of four velocity components.  We have numbered the line components from
[1] to [4] in order of increasing radial velocity.  Evidence in favour
of multi-peaked lines comes from the  appearance of [NII]\l6583
components [2], [3], [4] line in the region of {\bf B} and, in part,
{\bf A} (Fig. \ref{fig:blend}). It is likely that the subdivision into four
redshift systems
is an oversimplification.  Scans correponding to {\bf A} and {\bf D}
(Fig.  \ref{fig:blend})  suggest that another component displaced by
$v_r \sim 80$ \kms\ from [1] may also be present.

\subsubsection{Analysis of the H\a\ + [NII]\l\l 6548,6583 blend}

Although the blending is severe, multicomponent fitting with a single FWHM
value proved to be reliable (i. e., results did not depend upon details in the
placements of the continuum or
initial estimate of the peak positions).  We detail our procedure in the
following paragraph.

Component [1] (V$_r\approx$ 14250 \kms\ ) of [NII]\l6548 is not contaminated
by any other line, so  we constructed  a template [NII]\l 6583 [1]
using a Gaussian profile of the same width, taking advantage of the
fact that the intensity ratio [NII]\l 6548/[NII]\l 6583 is fixed.
Components [2], [3]. and [4] of [NII]\l 6583, although heavily blended,
are not contaminated by other lines. A deblending procedure was applied
to these three lines as a first step. After deblending of [NII]\l 6583,
templates were constructed for [NII]\l 6548 at the expected
wavelengths. [NII] emission was subtracted from the original spectrum,
leaving a {\em pure H\a} blend. A multi-gaussian deconvolution
procedure was then applied to the H\a\ blend as well. The
strength and width of the [1] and [4] components were found  to be
well constrained; the uncertainty due to the deblending procedure is
estimated to be $\approx$ 10 \%, and should be added to the uncertainty
in the flux calibration.  We report the decomposition results in Table
2 where we give flux, EW, and FWHM (only for [1] and [4]; on results are given
for {\bf F}, since
the S/N ratio is low) and the total flux
and EW in each line (the total flux of the four components) for all regions
where
emission lines show composite profiles.

\subsection{Emitting Regions at P. A. = 8$^\circ$}

{\bf A} -- The overall spectrum of {\bf A} is shown in Figure \ref{fig:spec}.
It
is more suggestive of FIR--loud galaxies or LINERs than of classical
Starburst and Seyfert galaxies. The width of the [OI]\l 6300 line is
for instance similar to the width in NGC 1068, the prototypical Seyfert
2 galaxy, but the ratio [OIII]\l5007/H\b\ $\sim$ 1 is too low for a
Seyfert classification (Osterbrock 1978).  The H\a\ + [NII]\l\l6548,6583
blend appears smoother than in {\bf B} and blending
of the various components is more severe in this region allowing almost
no discrimination of the suspected multiple components.  The FWHM $\sim
400$ \kms\ of each component is somewhat larger here than in the other
regions.

The radial velocity of the underlying galaxy estimated from NaD is $v_r
\approx$  14,200 \kms. The velocity curve at 0" (P.A. = 116$^\circ$)
gives a consistent value, $v_r \approx$ 14,200 \kms. These values are
also consistent with the HI profile reported by Bothun et al.(1983),
from which we estimated a systemic radial velocity $\approx$ 14,198
\kms. The RC3 reports $v_r(HI) \approx$ 14,265 \kms, but the HI profile
of {\bf S} is disrupted on the long--wavelength side, making a mean
velocity uncertain. The multiple line components show $v_r$ $\approx$
14,250 $[$1$]$, 14,700 $[$2$]$, 15,000  $[$3$]$, 15,450 $[$4$]$ \kms.
The lowest-velocity in the multiple
emission region is similar to the $v_r$\ of the underlying galaxy.
The continuum is rising toward the blue in F$_\lambda$,
suggesting an enhanced star-formation rate (\S \ref{sec:sfr}).

{\bf B} -- The emission line luminosity of {\bf B} is comparable to
that observed in {\bf A}, but the underlying continuum is significantly
weaker. We detected no absorption lines.  The emission line ratios also
differ from those of {\bf A}. Both components [1] and [4] show [N
II]/H$\alpha$ consistent with stellar photoionization, but the strength
of [O I] $\lambda 6300$, especially in [4], may require some additional
source such as shocks. Radial velocities of the four components are:
 $v_r$ $\approx$
14,350 $[$1$]$, 14,690 $[$2$]$, 14,960  $[$3$]$, 15,370 $[$4$]$ \kms.

{\bf C} -- Emission line profiles are single--peaked and barely
resolved in our spectra. Emission line ratios are typical of HII
regions.

{\bf D} -- This is a site of strong off--nuclear line emission.  The
ratio [NII]\l6583/H\a\ $\sim$\ 1.5 for [1] is larger than in {\bf A}.
The redshifted line components are weaker compared to [1] (e.g.
H\a\ luminosity of [4] is only $\sim$ 0.5 that of [1]) than seen at
{\bf A}. The maximum value of
the ratio [NII]\l6583/H\a\ in component [1] is seen at {\bf D}. On {\bf D}
the redshifted emission fades rapidly, giving the appearance
of a broad wing to the lines (see \ref{fig:blend}). Line components $v_r$\ are
14,320 $[$1$]$, 14,870 $[$2$]$, 15,130  $[$3$]$, 15,380 $[$4$]$ \kms.

{\bf CPG 29 N} $\equiv$ {\bf E} -- We derive $v_{r,E} =
15,000 \pm 30$ \kms from the NaD line for the resessional velocity of {\bf E},
in agreement with  Tifft
(1982). Karachentsev, Sargent \&\ Zimmermann (1979) give $v_{r, E}
\approx 14,806  $ \kms, while Karachentsev (1980) gives 14,429 \kms.
We note that the redshift given in Karachentsev (1987) $v_{r, E} \approx
14569$ \kms\ is a typographical error.  Mazzarella \&\ Boroson
(1993) recently report a surprising $v_{r,E} \approx $ 14,553\kms.  A
check of the frame that contains both {\bf E} and {\bf S} (P. A. =
8$^\circ$), confirms a shift in the two spectra at NaD that is
consistent with a velocity difference $\Delta v_{r,E-S} \simgt 700$
\kms.

\subsection{Velocity Curves}

Velocity curves from our long-slit spectra are given in Figs. \ref{fig:pa8},
\ref{fig:pa116}, \ref{fig:pa81}, along with 2D contour
displays of the data. The spatial zero points are set from the
continuum peaks of the nucleus ({\bf A}) for Figs. \ref{fig:pa8} and
\ref{fig:pa116}, and from knot {\bf B} for Fig. \ref{fig:pa81}.  We will note
here some of the most important features of the velocity structure in each
case.

PA $116^\circ$ -- This is the major axis of {\bf S}. The western part
of the disk has a plausible rotation curve for a typical spiral, rising
rapidly from the nucleus. The central velocity may be affected by the
presence of systems [2] and [3]; [4] is well seen near the nucleus
proper and is not plotted in the velocity slice. The contour
representation shows that the  peak emission in [4] is displaced by
about 3 arcseconds west of the continuum peak of the nucleus {\bf A}.

PA $8^\circ$ -- The slit passed through the nuclei of both {\bf S} and
{\bf E}. Only {\bf S} is shown in Fig. \ref{fig:pa8}.  The anomalous
emission-line systems appear to best advantage in this orientation, and
systems [1] and [4] can be traced in detail. Velocity gradients are
resolved in both systems, and at this position angle are almost along
the projected minor axis of {\bf S}. This may be evidence for
noncircular motion in the disk gas, since simple deprojection gives
implausibly large velocities within the disk plane.

PA $81^\circ$ -- This observation includes knot {\bf B} and is roughly tangent
to the NW spiral arm. The anomalous systems are confined to
a small region around {\bf B}, with the spiral arm showing small-amplitude
structure not unusual for large spirals.

\section{Direct Implications}

\subsection{Ionization Mechanisms}
\label{sec:ion}

Figure \ref{fig:ion} presents diagnostic diagrams, according to the
prescription of Osterbrock \&\ Veilleux (1987), for several emission
line regions.  Since we lack the ratio [OIII]\l5007/H\b\ for some
regions, we also show a combination of diagrams based on emission line
ratios detected in the red only.  The thick solid curve tracks a
semi--empirical dividing line between thermal (HII regions photoionized
by hot stars) and non--thermal (shock heated  or photoionized by a
power-law spectrum) ionization mechanisms (Osterbrock 1989).  Generally
speaking, all single-component emitting regions show line ratios
consistent with photoionization by stars, although several regions are
located in a borderline position because of their rather high
[SII]\l\l6717, 6730/H\a\ ratio.

The analysis is more complex for the multi-component lines where the
dominant ionization source may be non-stellar. The ratio
[OIII]\l5007/H\b\ $\simlt$ 1 indicates an ionization level too low for
classical AGN, in fact, {\bf A}, {\bf B}, and {\bf D} have almost LINER spectra
with [SII]\l\l6716,6731/H\a, and [OI]\l6300/H\a\ too high to be due to HII
regions, but with component [4] closer to the ratios typical of HII regions
than component [1], at least on {\bf A} and {\bf B} (Table 2).
Photoionization by a hidden AGN cannot be responsible for the
extended non-thermal emission observed at  P. A. = 8$^\circ$ without
violating the observed line intensities, since maintaining nearly
constant $\Gamma$ would require a corresponding drop in plasma density
and, therefore, in emitted surface brightness between the nucleus and
off-nuclear regions, which is not seen.  A compact source of ionizing
photons, like an AGN, would also produce a gradient in excitation
degree and line luminosity as a function of radius in the extended
emitting region. The ratio [OIII]\l5007/H\b\ $\sim 1$ occurs both at
{\bf A} and in the surrounding region as far as  10" radius.  We do
{\it not} observe very high ionization gas ([OIII]\l5007/H\b\ $\gg 1$)
anywhere in the galaxy.  Both findings suggest that the ionization
source is most probably provided {\it in situ}.  Photoionization by hot
stars in a dense medium has been proposed to explain the occurrence of
LINERs  in early type galaxies, where the interstellar pressure can be
high (Shields 1992). Although this may be the case for {\bf A}, it is
not easy to invoke densities as high as $n_e \sim$ 10$^{5.5}$
cm$^{-3}$\ in a more extended region in a spiral galaxy.  Shock
heating therefore seems the most plausible mechanism for the extended emission
at
P.  A. = 8$^\circ$ (which samples most of the regions of multiple line
components).

A comparison between the observed and predicted line ratios computed by
models of radiative shocks (Binette, Dopita, \&\ Tuohy 1985) shows that
several components have ratios close to the values expected for gas
heated by a slow shock $v_S \approx$ 86 \kms, with post--shock
temperature $T_e \approx 10^{5.2}~^\circ$K, and maximum post--shock
density of 170 cm$^{-3}$ (model B52 of Binette et al. and model D of
Shull \&\ Mc Kee 1979). Under these conditions, the predicted line
ratios are [OIII]\l5007$\sim$1, [NII]\l6583/H\a $\sim$ 0.65, [SII]\l\l
6716,6713/H\a\ $\sim 1.33$, and [OI]\l6300/H\a\ $\sim$ 0.44 which  are
rather close to the observed values. Shocks velocities up to $ \sim 200$ \kms
give plausible line ratios.
Hence moderate velocity  shocks could be the sole ionizing source.

\subsection{The Galactic Masses}

The mass of {\bf E} is probably equal to or greater than that of {\bf
S}. Assuming a mass--to--light ratio M/L$_B$ = 8 for {\bf E} and 3 for
{\bf S} (e. g., Faber \&\ Gallagher 1979), we estimate from the face-on
luminosity  M$_{Spiral} \approx 2.1 \times 10^{11}$ h$^{-2}$ M$_\odot$, and
M$_{Ell} \approx
2.5 \times 10^{11} $h$^{-2}$ M$_\odot$. Assuming that the $v_r$\ curve
on the western side of the galaxy is due to rotational
motion, we obtain M$_{Spiral} \sim 1.1\times
10^{11}$h$^{-2}$ M$_\odot$\ within d$\approx 17.2$ \h\ kpc. The
dynamical mass of the elliptical is M$\approx 2R <\sigma^2>/$G $\approx$
2.5$\times 10^{11}$ h$^{-2}$ M$_\odot$. Given the uncertainties in the
mass estimates (about a factor 2), we assume that the masses of {\bf E}
and {\bf S} are comparable. The line of sight component velocity
difference of $\sim$750 \kms suggests that the two galaxies form a
bound system only if the total mass  is M$\simgt 1.2\times
10^{12}$M$_\odot$, or  a factor $\simgt$ 5 larger than the value
estimated above. We note in passing that a bound system including CPG
29 and Mrk 983 would have mass $M_{AllBound} \simgt
1.60\times10^{12}$h$^{-1}$ M$_\odot$. This value is not implausible, if
compared to the dark matter amount estimated  for other groups of
galaxies.

 \subsection{Clues to the Star Formation History}
\label{sec:sfr}

The optical continuum of {\bf A} rises to the blue to our
spectral limit of $\sim 4,400$ \AA. Such an increase is not usually
observed in spiral galaxies of type Sc and luminosity class I (although
blue spectra are frequently observed in the luminosity classes  from II
to V, Bica \&\ Alloin 1987). We ascribe the blue spectral excess (see
Fig. \ref{fig:spec}) to an enhancement in the young stellar population
(e. g. a starburst component).
We subtracted the  continuum spectrum of a typical Sc I galaxy
(NGC 692) from the spectrum of {\bf A} in order to isolate the
starburst component (the dotted line in Fig. \ref{fig:spec}
shows the {\em adopted} continuum of the old stellar component).  Only a few
heuristic considerations are possible from our
data.

The blue continuum excess and the presence of strong  absorption
underlying the H\b\ emission line suggest that the integrated spectral
type of the starburst component is earlier than A5. Assuming that only
A0 stars contribute to the Starburst spectrum at $\lambda \sim 4800$
\AA, $\sim 10^8$ A0 stars are needed to produce the  continuum
luminosity of the Starburst component.  Given an Initial Mass Function
$\Psi(m) \propto m^{-\alpha}$, with \a\ $\approx$ 2.5, the number of
O--B stars produced (10--50 M$_\odot$) is n(O--B) $\approx$ 0.13 n(A).
The duration of the starburst must be much longer than the main sequence
lifetime of O--B stars $\tau_{ms,OB} \sim 5\times 10^6$ yr; otherwise
the H\a\ luminosity L$(H\alpha)_{SF}$ would be a factor 10$^2$ larger
than observed, assuming a gas covering factor $f_c\approx 0.1$.
We can infer that the SFR must have faded in the last 5$\times10^6$yr, and that
the duration
of the Starburst might be not longer than the main sequence lifetime of A0
stars
$\tau_{A0} \sim 10^8$ yr. The total stellar mass contributed by the Starburst
in $\sim 10^8$yr would be $\simgt 10^9$\msol.



Region {\bf R} is a site of strong off-nuclear star formation. The
H\a\ luminosity in the western line emitting regions suggests a SFR of
$\approx$ 1.5 M$\odot$ yr$^{-1}$ adopting the IMF used above  (e. g.,
Bushouse 1987).

\section{Discussion}
\label{sec:disc}
The challenge in understanding CPG 29 centers on explaining the
velocity excesses up to 1300 km s$^{-1}$ that the emission-line
components north of the nucleus show relative to the redshift of the
{\bf S} nucleus. We explore four possible interpretations for the
peculiarities observed in {\bf S} including both interaction and
internal mechanisms.

\subsection{Interpretation 1: The Spiral as an FIR Strong Galaxy}

Large internal velocity ranges with split emission line profiles have
been observed in the most powerful IRAS (hereafter FIR Strong)
galaxies.  The degree of complexity observed in the blend
H\a+[NII]\l\l6548,6583 is comparable to that observed in NGC 3079
(Heckman, Armus \&\ Miley 1990) and in IRAS 19254--7245 (Colina, Lipari
\&\ Macchetto 1991). The diagnostic emission line ratios are similar to
the ones observed in strong FIR emitters (e. g., Heckman, Armus
\&\ Miley, 1990).

The FIR luminosity of {\bf S} derived from  IRAS ADDSCAN data is
L$_{FIR}$ $\approx 2.6 \times 10^{10}$h$^{-2}$ L$_\odot$ (we neglect
any contribution from the elliptical component which is likely to be
negligible). This is about 8$\times$ the average for the ES pairs
listed in the CPG (Xu and Sulentic 1991). This value is also larger
than those observed for the prototypical starburst galaxies NGC 253 and
M82. The inclination corrected  magnitude for {\bf S} is $m_{0T}
\approx $ 14.20 (RC3) which  corresponds to L$_B \approx 5\times
10^{9}$ h$^{-2}$ L$_\odot$.  The ratio L$_{FIR}$/L$_B$\ is therefore
$\approx$ 4.6, again larger than the value for NGC 253, but lower than
the values observed in the most powerful FIR galaxies. The color ratio
$f_{60\mu m}/f_{100\mu m} \sim 0.30$ implies a dust temperature of
$\approx$ 47 $^\circ$K, lower than that observed in most FIR galaxies.
The values of L$_{FIR}$, of $f_{60\mu m}/f_{100\mu m}$, and of
L$_{FIR}$/L$_B$\ place Mrk 984 at the low--power end of the strong FIR
emitters studied by Heckman et al.  (1990). The low  $f_{60\mu m}/f_{100\mu
m}$\ is not atypical of late type normal spirals.

Large-scale bipolar outflows along the disk rotation axis, as found in
FIR strong galaxies such as NGC 253, 3628, and 6240, would give rise to
both blue and red--shifted emission line components. The opacity of the
disk could suppress the blue--shifted emission if the outflow were
close to the disk plane and the disk were properly warped.  The
extinction in the disk of the galaxy should be A$_V \simgt$ 4.5 mag to
dim the blue--shifted H\a\ emission down to the $3\sigma$ level (under
the assumption that the red--shifted  emission line components are not
obscured). White \&\ Keel (1993) and James \&\ Puxley (1993) recently
estimated that the disk opacity of face--on Sb--Sc galaxies should only
give $A_V \simlt 1.8$ mag. The appearance and inferred inclination  of
{\bf S} makes a value as large as 4.5 a not unlikely possibility.
However, the total $\Delta v_r$\ (i. e., including the obscured
blueshifted component)  should be $\approx$ 2,600 \kms, which exceeds
the value observed in all of the most powerful FIR galaxies.  The
dimming of the blue--shifted component of an outflow would require a
conspiracy of two extreme conditions - outflow close to the disk plane
and warping of the disk so that only the redshifted outflow remains
unobscured.

\subsection{Interpretation 2: Interaction with the Intergalactic Medium}

This is a relatively old idea (e. g., Freeman \&\ de Vaucouleurs 1974)
more recently discussed by Combes et al. (1988) in a context similar to
CPG 29. It was recently revived by the discovery of diffuse X-ray
emission in the NGC 2276/2300 ``group'' of galaxies (Mulchaey et al.
1993) which is dominated by the relatively isolated E+S pair CPG 127
(Arp 114). One could hypothesize that {\bf S} is passing through a
region where the ram pressure of a relatively dense intergalactic
medium is stimulating star formation and shocks. The compressed
appearance of the bright arm on the northern edge of {\bf S} is
suggestive of such an effect. Mulchaey et al. (1993) also suggest that
the peculiar optical morphology of NGC 2276 is due to an ongoing
intragroup medium--galaxy encounter. This interpretation must remain
speculative until HI or X-ray data become available for the
neighborhood of CPG 29.

\subsection{Interpretation 3: Interaction Involving a Third Party}
\label{sec:ring}
\label{sec:3party}

While we cannot rule out the possibility of a third galaxy, it appears to be
well concealed. An observation possibly favoring the existence of a third small
galaxy, close to {\bf S} is that emission line ratios in {\bf C} and component
[4] of {\bf B} are consistent with HII regions. {\bf B} might be
identified with the small companion. However, component {\bf C} appears too
diffuse while {\bf B}
shows at best a weak continuum.  The HI data for CPG29 (Bothun et al. 1983)
also shows an emission profile consistent with a single gas rich galaxy.

\subsection{Interpretation 4: Pole--on Collision Between Spiral \&\ Elliptical}

\subsubsection{Morphology: An Emerging Ring Galaxy?}

Two ring--like features are easily visible on the CCD image of {\bf S} (Fig.
1, and \S \ref{sec:morph}). These features  can be produced by a  polar
or nearly polar encounter, in principle either with the elliptical, or
with a hypothetical third party (e. g., Toomre \&\ Toomre 1972; Theys
\&\ Spiegel 1977; Appleton \&\ Struck--Marcell 1987).
 Numerical simulations and physical intuition suggest that a primary ring
should
start forming at the time the intruder crosses the target galaxy;
assuming that the plane of the orbit is inclined $i\approx
45^\circ$\ to the line of sight, the separation time between spiral and
elliptical and the order--of--magnitude expansion times (in years) for the
rings
are:

\begin{tabular}{cccc}
 $\tau_{Ell,Spiral}$ & $\tau_{in}$ & $\tau_{inter}$ & $\tau_{outer}$ \\
$5.2\times10^7 \Delta v_{Ell-Spiral,1000}^{-1}$ &
$4.0\times10^7 v_{in,100}^{-1}$ & $9.4\times10^7 v_{inter,100}^{-1}$ &
$2.4\times10^8 v_{out,100}^{-1} $\\
\end{tabular}
\vspace{0.5cm}

The expansion velocity of the primary ring should be $\approx
\frac{1}{2} \Delta v_{Ell-Spiral}$; the outermost ``ring" i. e.,  the line  of
knots at the southern edge of the galaxy,  and a faint,  northeastern arc
should be associated with the
``aged" primary ring since $\tau_{out} \approx 4.8\times10^7 yr \approx
\tau_{Ell,Spiral}$.

Hydrodynamical models (Appleton \&\ Struck--Marcell 1987; Appleton
\&\ James 1990) suggest that the leading edge of the primary ring should be
a site of strong star formation. The secondary ring should show
even stronger star formation due to a large degree of compression in the
density wave; star formation should then fade with increasing ring order.
This scenario agrees with the star formation properties of the ring--like
features observed in {\bf S} (in the simplest picture of a single passage):
(1) no evidence of H\a\ emission in the innermost ring and (2) strongest star
formation in the bright arc of the intermediate ring. Little can be said
about the outermost line of knots that skirts the southern edge of {\bf S}
but it seems reasonable to suppose that they are HII regions.

The northern side of the intermediate ring (the bright arc, or {\bf R})
is much brighter than the southern one; this probably implies that the
encounter between the spiral and the elliptical was mildly
off--centered (impact parameter $\simlt 6"$). Gerber, Lamb \&\ Balsara
(1992) attempted to reproduce the morphology of Arp 147 through numerical
simulations of the the slightly off--center polar encounter of two
galaxies of equal mass. Their simulations produced  a feature that,
once projection effects are accounted for, can be equated to the {\em
bright arc} of {\bf S}.  In the  innermost regions of {\bf S} ($\simlt
6"$), the motion of the nucleus across the galactic disk may have
contributed to the creation of a surrounding shock heated region and,
even, to the double peaked appearance of the nuclear region in the
H\a\ image. We note that the morphology of the line emitting gas is
markedly different from that of the stellar component. Simulating separately
gaseous and stellar components in head-on encounters, Horellou \&\ Combes
(1993) showed that gas can be removed out of one of the colliders: gas would
accelerates at high velocities toward the intruder (after its passage),
creating a vertically displaced structure, while the stellar component does
not, in agreement with what we observe in CPG 29.

\subsubsection{Simulations: Could the Elliptical Emerge Unscathed?}
\label{sec:nbody}

Evidence for strong interaction in ellipticals often includes a displacement of
the outer isophotal centroid towards the companion; the side closest to
the companion is often distended, while the opposite side is compressed
(Borne 1988). From the analysis of \S\ \ref{sec:morph}, we can conclude
that {\bf E} shows little, if any, evidence for significant tidal
perturbation. Small deviations from an $r^{1/4}$\ law are frequently
observed in ellipticals, although their significance with reference to
the interaction history is not clear (Djorgovski \&\ Kormendy 1989).

We carried out  numerical simulations of the pair,
employing the N--Body ({\tt NBODY2}) code written by S. Aarseth. Our
aim was not to reproduce satisfactorily the observed morphology of {\bf
S}, but rather to study whether the elliptical can emerge unscathed by
the pole--on or nearly pole--on crossing of the spiral.

We used $\approx$ 1,200 particles for each galaxy in each simulation; we
modelled {\bf S} as  flat exponential disk with e--folding scale 1.5 -- 2.5
kpc; the elliptical was represented as a modified isothermal spheroid (King
radius 3 kpc) supported by anisotropic velocity dispersion. The stability of
both spiral and elliptical was studied in several control runs in which both
galaxies were  allowed to evolve as  isolated objects. We assumed that the two
galaxies have the same mass, $M \approx 2.5\times 10^{11}$ M$_\odot$. We
simulated various encounters for different choices of the inclination
(i=0$^\circ$, 15$^\circ$) and of the initial approaching velocity, and for two
values of the impact parameter (0" and 6").  For  $\Delta v_{Ell-Sp} \approx$
1,000 and 700 \kms\ the elliptical is not affected by the crossing of the
spiral. This result of our simulations has been confirmed by more sophisticated
simulations of head--on elliptical/spiral encounters performed employing the !
{\tt TREECODE} by L. Hernquist. T

We will suggest in the next section how a collision between {\bf E} and {\bf S}
can easily explain both moderate velocity shocks and large internal $\Delta
v_r$ in {\bf S}.

\subsubsection{The Shock Heated Gas}

The HII--like line ratios on {\bf C} can be explained if
the gas has been stripped and accelerated towards {\bf E} by {\bf E} itself.
The
acceleration of molecular gas clouds depends upon time, because of
the varying potential of the approaching, crossing and then receding
elliptical galaxy. The clouds accelerated to the highest velocity will
not collide with other clouds. If clouds accelerate to a velocity $v_1
> v_2$ at $t_1 > t_2$, then the velocity difference could make a
cloud--cloud collision possible at a moderate velocity $v_{12} \approx
v_1 - v_2 \sim 100$ \kms. Cloud--cloud collisions are therefore a
likely agent in the production of shock heated gas for line components
[1]--[3] in {\bf B} and possibly {\bf A}. If the thickness of the
clouds is $\Delta s \approx 2.7\times 10^{20}$ cm, the clouds are
destroyed in a time t $\approx 8.6\times10^5 v^{-1}_{12,100}$\ yr. Shock
production is thus sustained for an appreciable amount of time. The
$\Delta v_r \sim$ 1,300 \kms in {\bf S}, is much larger than the radial
velocity difference between the components (750 \kms).  This does not demand
that
a third galaxy is responsible for the disk collision;  it rather demands that
part of the orbital energy be transferred to the stripped molecular clouds.

Models with shocks velocity higher than 250 \kms\ do not agree with the line
ratios observed in {\bf S}. However, shock--heating emission may be diluted by
photoionization by OB stars; in which case more extreme shock velocities could
be appropriate. Caution is needed in converting between observed line
velocities and expected shock velocities, since the latter are strictly local
and may, for example, reflect shocks driven into individual clouds rather than
large scale phenomena characterized by the large velocity splitting between
emission line components.

Binette, Dopita \&\ Tuohy's (1985) models give an
H$\alpha$\ luminosity per unit shock transversal area L(H\a)/$\Sigma
\approx 3.9 \times 10^{-5} v_{100}^{1.48}$ ergs s$^{-1}$ cm$^{-2}$. A typical
giant molecular cloud in the solar neighborhood has mean diameter $\approx$
45 pc, or a projected surface area $\Sigma \approx 2.1\times 10^3$ pc$^2$;
if two molecular clouds of this size collide with $\Delta v\approx 100$
\kms, we will have L(H\a) $\approx 2\times10^{40}$ ergs s$^{-1}\times
L(H\alpha)/\Sigma
\times (t_{cool}/t_{dyn}) \approx 1.7 10^{37} \times v_{100}^{4.94}$ ergs
s$^{-1}
\approx 4.4\times 10^3 v_{100}^{4.94}$ L$_\odot$, where the ratio t$_{cool}$\
is the cooling time, and t$_{dyn}$\ is the time needed by the shock front to
cover
a distance equal to the tickness of one cooling layer in the post-shock zone.
The L(H\a) of the
emitting regions where shock heating is the dominant
ionization mechanism is $\sim$ 1$\times 10^{41}$ h$^{-2}$ ergs
s$^{-1} \approx 2.6\times 10^7$ h$^{-2}$ L$_\odot$. The equivalent of
$\sim$ 10 giant molecular clouds must be presently colliding in {\bf S}
in order to produce the observed L(H\a) if $v \sim 200$ \kms. Assuming that
each cloud has a
thickness of 60 pc, and that the average density in the clouds is $\sim
100$ cm$^{-3}$, the mass of molecular gas needed to sustain the
H\a\ emission is M$_{H2} \approx 1.2 \times 10^7$ h$^{-2}$ M$_\odot$.

At face value, the observed L$_{FIR}$\  implies a current SFR of
13 M$_\odot$ yr$^{-1}$; however shock-heated gas may be responsible for an
enhancement of FIR luminosity as well (e. g., Harwit et al. 1987).
Equating the kinetic energy dissipated in the intra--cloud collision to
the L$_{FIR}$\ observed for {\bf S}, requires a mass M$ \approx
2.12\times10^{10} Z_{60} L_{10} v^{-3}_{100}$ M$_\odot$. The molecular
content of an entire galactic disk must be involved in the collision
unless the velocity of approach is much larger than 100 \kms: only a violent
collision with a hypothesized third party, with $\Delta v \approx 1000$ \kms
would change the estimate in favour of shock heating explanation for $L_{FIR}$.
If we ascribe the H\a\ emission to shock heated gas in a layer between
colliding molecular clouds stripped by {\bf E} then only a negligible fraction
$\ll 1$\ of
$L_{FIR}$ can be due to shock heated gas, and
L$_{FIR}$ can be safely ascribed to star formation alone.

\subsubsection{The Problems with the Pole--on Collisions}

There some difficulties for this simple scenario.
A preliminary analysis of the
velocity curves at P. A. = 81$^\circ$\ suggest that the velocity field
deduced from line component [1] could be entirely rotational on the
west side of {\bf S} (\S\ \ref{sec:spec}).  Results are qualitatively
in agreement with the double--wave model {\em only} if the systemic
radial velocity is $v_{r,sys} \approx 14,260$ \kms; otherwise a
double--wave fit is not possible. We do not observe evidence of expansion
velocities as
large as 500 \kms. As shown previously, radial velocity differences on
the western side of {\bf S} do not exceed 150 \kms.


\section{Conclusions \&\ Implications}
\label{sec:concl}

We have presented images and spectra of the mixed-morphology galaxy pair CPG
29. The most unusual feature is the complex extended line emission activity in
{\bf S}. Our resolution of the emission lines into
multiple components has provided clues to the nature of this system.
We have shown that the morphology of  {\bf S} can be explained if {\bf S} is a
ring galaxy
emerging from a near pole--on encounter with {\bf E}. Knot {\bf B} and {\bf C}
probably represent tidally stripped gas rather than a second nucleus or a
smaller, third galaxy.  The large internal velocity difference of 1300 \kms\
and the shocked gas conditions in {\bf S} probably result from  the
acceleration of the line emitting gas by gravitational forces.

The results for {\bf S} raise the possibility that the acceleration of gas in
superwind galaxies  might be gravitational in origin, instead of  being
produced by the collective effort of ``organized'' supernov\ae. A favourable
orientation for the orbit of the companion, along with multiple passages, might
produce a two--way streaming of gas. This point is speculative, but it should
be investigated before any conclusion on the origin of optical line emission in
strong FIR galaxies is reached. This is especially true since the most powerful
FIR galaxies are probably merging spirals.  For instance, {\bf S} is kin to
NGC1143, a FIR strong spiral galaxy with a Seyfert--2 nucleus  that has
probably been crossed by a less massive elliptical companion. We can identify
several similarities between the morphology of the two galaxies. A picture of
NGC 1144 has been published by Hippelein (1989, Fig. 1). This author noted that
{\em the internal velocity differences in NGC 1143 are extr!
emely high}, with $\Delta v_r\sim

Hernquist (1989) showed that strong interaction is able to channel $\sim 10^9$
M$_\odot$\ of gas  within the innermost d$\sim 200$ pc. The fate of the gas at
$d \simlt 200$ pc is still unclear.  Probably bar--like instabilities are then
able to funnel the gas toward the innermost pc
(Shlosman, Frank, \&\ Begelman 1989).  It is important to stress  that we
observe strong
emission line activity, but little or no evidence for non--thermal nuclear
activity.
Heuristic consideration suggest that an {\em aged} nuclear starburst may
have produced a total mass of $\sim 10^9$M$_\odot$. This finding may
suggest that the bar--like instability supposed to channel the gas to
d$\simlt 1$pc is governed by the global galactic potential, and does not
operate in galaxies of late morphological type (e. g., Noguchi 1993).

\acknowledgments

D. D.--H. \&\ P. M. acknowledge the financial support of the University of
Alabama, and the support of the SPM Observatory staff, that allowed one
successful observing run. We acknowlege support from EPSCoR grant EHR-9108761.
D. D.--H.  acknowledges support from grant IN451091 from DGAPA, UNAM.
Observations at the 6-meter telescope were supported by the NSF under
the Large Foreign Telescopes program. We thank Anna Curir for the simulation
performed.
\newpage

\vfill\eject\newpage

\textwidth 18.0 truecm
\textheight 23. truecm
\oddsidemargin -1.5 cm
\evensidemargin -1.5 cm

\tightenlines
\begin{table}
\small
\begin{center}
\caption{Journal of Observations}
\medskip
\begin{tabular}{ccccccccccc}
\tableline\tableline
Obs. & Tel. & Detector  & Date & U.T. & E.T. & Filter & Scale & Disp. & P. A. &
Sp. Range \\
& & & &  && & ("/pixel) & (\AA/mm) & & (\AA) \\
\tableline
KPNO   & 2.1m & Tek 512$\times$512 & 1986 Nov 29& 5$^h$59$^m$ & 20$^m$ &  B &
0.3429 &...&...&...\\
KPNO   & 2.1m &   Tek 512$\times$512 & 1986 Nov 29& 4$^h$16$^m$ & 15$^m$ & V &
0.3429 &...&...&...\\
Lowell & 1.1m & TI 800$\times$800 & 1991 Sept 12 & 8$^h$07$^m$ &  90$^m$ &
H$\alpha$ & 0.708 &...&...&...\\
Lowell & 1.1m & TI 800$\times$800  & 1991 Sept 12 & 9$^h$40$^m $ &  30$^m$ & R
& 0.708 &...&...&... \\
\\
KPNO & 2.1m  & TI 800$\times$800 & 1990 Nov 23 & 2$^h$17$^m$ & 45$^m$ & ...&
0.782 & 85   & 8$^\circ$ &
6140--7160 \cr
SPM  & 2.1m &   Thompson 385$\times$576 & 1991 Dec 6 & 5$^h$51$^m$ & 40$^m$ &
...& 0.873&114  & 7$^\circ$ &
6320--7450 \\
SPM  & 2.1m & Thompson 385$\times$576 &1991 Dec 6  & 6$^h$58$^m$ & 30$^m$
&...&0.873& 114  & 116$^\circ$ &
6320--7450 \\
ESO  & 1.5m &    FA2K &1992 Jul 1  & 9$^h$44$^m$ & 60$^m$\tablenotemark{a}
&...&0.68& 159   &
3$^\circ$\tablenotemark{b} & 3670--7580\\
ESO  & 1.5m &  FA2K& 1992 Jul 2  & 9$^h$07$^m$ &  80$^m$\tablenotemark{a}
&...& 0.68& 159  & 81$^\circ$\tablenotemark{c} &
4530--7580\\
KPNO & 2.1m & Ford 1024$\times$3072 & 1992 Niv 14 & 5$^h$46$^m$ & 60$^m$
&...&0.782 & 85  & 82$^\circ$ &
4470--8330\\
\tableline
\end{tabular}
\end{center}
\tablenotetext{a}{Sum of two consecutive exposures}
\tablenotetext{b}{Focusing problem; extracted only regions {\bf A} and {\bf B}}
\tablenotetext{c}{Focusing problem; used only for confirmatory
purposes}\label{tbl-1}
\end{table}
\newpage

\voffset=-1.1cm
\begin{table}
\begin{center}
\small
\caption{{\bf S}$\equiv$ Mrk 984: Emission Lines Fluxes \&\ Widths of Regions
with Multiple Line Profiles} \medskip
\begin{tabular}{llcccccccccc}
\tableline\tableline
Region & Ident. & \multicolumn{2}{c}{Total} & & \multicolumn{3}{c}{[1]} &  &
\multicolumn{3}{c}{[4]}  \\
\cline{3-4} \cline{6-8} \cline{10-12}
& & Flux\tablenotemark{a} & EW\tablenotemark{a} &
& Flux & EW & FWHM\tablenotemark{a} && Flux & EW & FWHM \\
\tableline
{\bf A}\tablenotemark{b} \\
& $[$SII$]\lambda\lambda$6716,6731 & 25 & 39&  & 8.4 & 14 & 380 && ... & ...
& 380 \\
&$[$NII$]\lambda$6583    &   22.4 &  32&  &  10.3 & 15  & 370 &&  4.02 & 6 &
370\\
&H$\alpha$               &  27.0  &  40 & &  9.6 & 14 & 400 &&  5.73 & 8.5 &
400\\
&$[$NII$]\lambda$6548    &   7.6  & 11  & &  3.4 & 5 & 370 & & 1.34 & 2 & 370\\
&$[$OI$]\lambda$6363     &   $<$2.8  & $<$4 &  &...  & ...  & ... &
&  ... & ... & ... \\
&$[$OI$]\lambda$6300     &   6.2  & 10 &&  3.4  & 2.5  & 410 & & 1.87 & 2.2 &
410 \\
%
{\bf A}\tablenotemark{c} \\
&$[$SII$]$$\lambda\lambda$6716,6731 & 14.5 & 62 && ... & ... & ... &&
... & ... & ... \\
&$[$NII$]$$\lambda$6583 & 14.4 & 62 && 8.46 & 36 & $\simlt$220 && $<$2.27 & 9.7
& $\simlt$220  \\
&H$\alpha$  & 8.57 & 32 && 2.93 & 12.5 & $\simlt$220 && 3.36 & 14.5 &
$\simlt$220 \\
&$[$NII$]$$\lambda$6548 &  4.89 & 21 && 2.87 & 12.2 & $\simlt$220
&& 0.76 & 3.3 &
$\simlt$220 \\
&$[$OI$]$$\lambda$6300  & 4.68 & 20 && 0.48 & 4 & 280 && 1.59 & 6.5 & 280 \\
%
{\bf B}\tablenotemark{b} \\
&$[$SII$]$$\lambda\lambda$6716,6731 & 18.4 & 52 && ... & ... & ... && ... &
.. & ... \\
&$[$NII$]\lambda$6583   & 14.7 & 45 && 2.45 & ... & $>$180 &&
5.1 & 16 & 340 \\
&H$\alpha$              & 29.6 & 94 && 5.31 & 17 & 240 && 12.8 & 40 & 240 \\
&$[$NII$]\lambda$6548   & 5.0 & 15 && 0.83 & ... & 180 &&  1.73  & 5  & 340\\
&$[$OI$]$$\lambda$6300  &  4.4 & 12  &&  0.42 & 1  & 240 && 1.52 & 4 & 240 \\
{\bf B}\tablenotemark{d} \\
&$[$SII$]$$\lambda\lambda$6716,6731 & 21.0 & 57 && 8.34 & 22 & 270 && ... &
.. & 270 \\
&$[$NII$]$$\lambda$6583     & 18.5  & 39 && 6.50 & 18 & 250 && 5.34 & 14 &
340 \\
&H$\alpha$ & 32.0 &   81     && 11.8 & 30  & 210 && 10.7 & 27 &
250 \\
&$[$NII$]$$\lambda$6548   &  6.2  & 13   && 2.5  & 6  & 250 && 1.81 & 5& 340\\
&$[$OI$]$$\lambda$6363 & $<$2 & $<$5 && ... & ... & ... && ... & ... & ... \\
&$[$OI$]$$\lambda$6300
&   4.95 & 14   && ...  &  ... & ... && ... & ... & ... \\
&$[$OIII$]$$\lambda$5007    & 12.36 & 35 &&  4.20 & 12 & 375 && 4.85 & 14 &
385  \\
&H$\beta$ &  11.93 & 33 && 4.35 & 12 & 330 && 4.97 & 14 & 385 \\
{\bf D}\tablenotemark{b}\\
&$[$SII$]$$\lambda\lambda$6716,6731 & 22.2 & 47 && ... & ... & ... && ... &
.. & ... \\
&$[$NII$]$$\lambda$6583 & 21.0 & 45 &  &  13.8 & 30 & 375  && 2.15 &  5  & 310
\\
&H$\alpha$              & 17.8 & 40  & &  9.76  & 22 & 425  && 4.29 & 10 & 430
\\
&$[$NII$]$$\lambda$6548  & 7.2 & 16   &&   4.7 &  10&  375&&  0.75 & 1.8 & 360
\\
&$[$OI$]$$\lambda$6363 &  ... & ...    && ...  &  ... & ... && ... & ... & ...
\\
&$[$OI$]$$\lambda$6300 
&  5.0 & 12   &  & 2.5  &  6 & 360 && 0.94 & 2.5 & 360 \\
{\bf F}\tablenotemark{c} \\
&$[$SII$]$$\lambda\lambda$6716,6731 & 4.53 & 41 && ... & ... & ... &&
... & ...  & ...\\
&$[$SII$]$$\lambda$6731 & ... & ... & &... & ... & ... && ... & ... & ... \\
&$[$SII$]$$\lambda$6716 & ... & ... & &... & ... & ... && ... & ... & ... \\
&$[$NII$]$$\lambda$6583 & 5.89 & 53 & &... & ... & ... && ... & ... & ...  \\
&H$\alpha$  & 5.25 & 48 & ... & ... & ... && ... & ... & ... \\
&$[$NII$]$$\lambda$6548 &  2.00 & 18 & & ... & ... & ... && ...  & ... & ... \\
&$[$OI$]$$\lambda$6300  & 1.91 & 17 && ... & ... & ...&& ... & ... & ... \\
\tableline
\end{tabular}
\end{center}
\tablenotetext{a}{Fluxes are in units of 10$^{-15}$ ergs cm$^{-2}$
s$^{-1}$; EW are given in \AA\ and FWHM in \kms}
\tablenotetext{b}{At P. A. = 8$^\circ$}
\tablenotetext{c}{At P. A. = 116$^\circ$}
\tablenotetext{d}{At P. A. = 81$^\circ$}
\end{table}
\newpage

\begin{table}
\small
\begin{center}
\caption{{\bf S} $\equiv$Mrk 984: Emission Line Fluxes \&\ Widths of
Regions with Single Peaked Emission Line Profiles} \medskip
\begin{tabular}{lccccccccccccccc}
\tableline\tableline
Ident. & \multicolumn{3}{c}{{\bf I}\tablenotemark{c}} & &
\multicolumn{3}{c}{{\bf R}
\tablenotemark{b}} &  &
\multicolumn{3}{c}{{\bf R}$\equiv$B. ARC\tablenotemark{c}}& &
\multicolumn{3}{c}{{\bf C}\tablenotemark{d}}  \\
\cline{2-4} \cline{6-8} \cline{10-12} \cline{14-16}
 & Flux\tablenotemark{a} & EW & FWHM && Flux & EW & FWHM && Flux & EW & FWHM &&
Flux & EW & FWHM \\
\tableline
$[$SII$]$$\lambda$6731 &  1.81 & 11 & $\simlt$170 &&  2.01& 15& 240 & &
                          2.24 & 13 & $\simlt$170 && 1.53 & 21 & 270  \\
$[$SII$]$$\lambda$6716 &  2.20 & 14 & $\simlt$170 &&  1.59 & 12 & 230 &&
                          3.00 & 17 & $\simlt$170 &&   2.06 & 27 & 270 \\
$[$NII$]$$\lambda$6583 &  4.02 & 24 & $\simlt$180 &&  3.21 & 26 & $\simlt$220
&&
                          4.58 & 28 & $\simlt$180 &&   1.81 & 23 &
$\simlt$220 \\
H$\alpha$              &  9.25 & 61 & $\simlt$180 &&  6.06 & 49 &
$\simlt$220 &&
                          13.1 & 81 & $\simlt$180 &&   6.35 & 65 &
$\simlt$220 \\
$[$NII$]$$\lambda$6548 &  1.67 & 12 & $\simlt$180 && 1.10 & 9 & $\simlt$220 &&
                          1.82 & 11 & $\simlt$180 && ... & ... & ...\\
$[$OI$]$$\lambda$6300  & $<$0.7 & $<$ 5 & ... & & 0.46 & 3 & ... &&
                          0.4   & 2     &  $\simlt$180  && 1.27 & 16 & 420: \\
$[$OIII$]$$\lambda$5007&  1.12  & 8 & 280 && ... & ... & ...  &&
                          1.30  & 5.5  & 215 && ... & ... & ...  \\
H$\beta$               &  1.75  & 9 & 205 && ... & ... & ... &&
                          4.64  & 23& 265 && ... & ... & ...  \\
\\
\\
Ident. & \multicolumn{3}{c}{{\bf J}\tablenotemark{c}} &&
\multicolumn{3}{c}{{\bf K}\tablenotemark{c}} &
& \multicolumn{3}{c}{{\bf M}\tablenotemark{b}} &&
\multicolumn{3}{c}{{\bf T}\tablenotemark{b}}  \\
\cline{2-4} \cline{6-8} \cline{10-12} \cline{14-16}
& Flux & EW & FWHM  && Flux & EW & FWHM &&
Flux & EW & FWHM && Flux & EW & FWHM \\
\\
$[$SII$]$$\lambda$6731 &1.23  &12  &$\simlt$170 && 0.63 & 7 & 175 &
 & ... & ... & ... && 0.23 & 4.5&
$\simlt$ 220 \\
$[$SII$]$$\lambda$6716 & 1.61 & 16 & $\simlt$170 && 0.98 & 11& 175 &&
.. & ... & ... && 0.25 & 4.1 &
$\simlt$220 \\
$[$NII$]$$\lambda$6583 & 2.82 & 30 & $\simlt$180 && 1.33 & 16 & 170 &
& $<$1.91 & 7 & 265 &&  0.71 & 17
& 320:\\
H$\alpha$ & 7.24 & 73 & $\simlt$180 && 3.97 &
48 & 170 && 2.95 & 12  & $\simlt$220  &&  2.03 & 49 & $\simlt$220 \\
$[$NII$]$$\lambda$6548 & 0.96 & 9 & $\simlt$180 && ... & ... & ... &
& ... & ... & ... && ... & ... &  ... \\
$[$OI$]$$\lambda$6300  & 0.3 & 3 & $\simlt$180 && ... & ... & ... &&
 ... & ... & ... &&  ... & ... & ... \\
$[$OIII$]$$\lambda$5007& 0.66 & 5 & $\simlt$220 && 0.95 & 7.5 & 190 && ...& ...
& ...
&& ... & ... & ...  \\
H$\beta$&  1.80 & 13 & 210 && 1.13 & 9 & $\simlt$220&& ... & ... & ... && ... &
... &
..  \\
\tableline
\end{tabular}
\end{center}
\tablenotetext{a}{Fluxes are in units of 10$^{-15}$ ergs cm$^{-2}$
s$^{-1}$}
\tablenotetext{b}{P. A. = 116$^\circ$}
\tablenotetext{c}{P. A. = 81$^\circ$}
\tablenotetext{d}{P. A. = 8$^\circ$}
\label{tbl-6}
\end{table}

\newpage
\begin{figure}\caption[fig:vimage]{
V image of CPG 29, obtained by N. Sharp. North is to the top, East
is to the left. The ticks identify the three slit configurations used to obtain
the spectra. Location of regions {\bf A}, {\bf B}, {\bf C}, {\bf D} are also
marked.
\label{fig:vimage}}
\end{figure}

\begin{figure}
\caption[fig:contour]{Contour plots of the inner regions of Mrk 984$\equiv${\bf
S}, for the B (a) and
H$\alpha$ (b) images. North is to the top, East is to the
left.\label{fig:contour}}
\end{figure}

\begin{figure} \caption[fig:ell]{Geometrical properties of the elliptical
galaxy, CPG 29N. Left: V image, Right: B image. Top panel: P. A. of the major
isophotal
axis as a function of radius (in arcsecs); Middle panel:
Ellipticity; Bottom panel: surface
brightness (in arbitrary units) versus r$^{1/4}$.\label{fig:ell}}
\end{figure}

\begin{figure}
\caption[fig:pa8]{
Isophotal contours of the H$\alpha$ + [NII]\l\l 6548,6583 emission lines of
{\bf S} at P.A. = 7$^\circ$. Horizontal scale is arcsecs along the slit,
vertical scale is wavelength in \AA. The origin of the horizontal scale
has been set at the centroid of the continuum at $\approx$ 6450 \AA\ (rest
frame). Note the off-centering of the emission line blend with respect to
the continuum. Radial velocity curves. Filled circles: Redshift system [1]
(see text); filled squares: redshift system [4]. Vertical scale is
heliocentric radial velocity in km s$^{-1}$. \label{fig:pa8}}
\end{figure}

\begin{figure}
\caption[fig:pa116]{
Isophotal contours of the H$\alpha$ + [NII]\l\l 6548,6583 extended emission
of {\bf S} at P.A. = 116$^\circ$. Horizontal scale is arcsecs along the
slit, vertical scale is wavelength in \AA. The origin of the horizontal
scale has been set at the centroid of the continuum at $\approx$ 6450 \AA\
(rest frame). Note the off--centering of the emission line blend with
respect to the continuum. Radial velocity curve; Vertical scale is
heliocentric radial velocity in km s$^{-1}$. \label{fig:pa116}}
\end{figure}

\begin{figure}
\caption[fig:pa81]{
Isophotal contours of the H$\alpha$ + [NII]\l\l 6548,6583 extended emission
at P.A. = 81$^\circ$, off-centered by approximately 5".6 from the nucleus
of {\bf S} (e. g., with the slit centered on the brightest blob [{\bf B}]
and roughly aligned with the bright arc labeled as {\bf R}).  Horizontal
scale are arcsecs along the slit, vertical scale is wavelength in \AA. The
origin of the horizontal scale has been set at the centroid of the
continuum at $\approx$ 6450 \AA\ (rest frame). \label{fig:pa81}}
\end{figure}

\begin{figure}
\caption[fig:blend]{Extracted spectra of the H\a+[NII]\l\l6548,6583 blend along
the
slit at P. A. = 8$^\circ$. Each spectrum corresponds to a spatial increment
of approximately 2".5. Three line components of [NII]\l6583 form
the ``boxy" blend to the red. The arrows indicate the wavelengths of the
[NII]\l6548 [1] and of the [NII]\l6583 [2], [3], [4] components.
\label{fig:blend}}
\end{figure}

\begin{figure}
\caption[fig:spec]{ Spectra of {\bf A} and {\bf B} Horizontal scale is
wavelength in \AA, vertical scale is specific flux. The dotted line in the
Spectrum of {\bf A} is the assumed continuum of the old stellar population.
\label{fig:spec}} \end{figure}

\begin{figure}
\caption[fig:ion]{Diagnostic diagrams for the emitting regions and line
components [1]
and [4] of {\bf S}. Emission-line ratios are plotted in logarithmic scales.
Thick solid lines mark the boundary between HII--regions (thermal) and
non-stellar ionized nebul\ae\ (shock heated or NLR-like); the dotted line is a
$+2\sigma$\
boundary. \label{fig:ion}}
\end{figure}

\end{flushleft}

\end{document}